\documentclass[10pt,a4paper,twoside]{article}
\usepackage{epsfig}
\usepackage{baltlat6}
\usepackage{array}
\usepackage{here}
\usepackage{wrapfig}
\begin{document}
\ \
\vspace{0.5mm}
\setcounter{page}{1}
\vspace{8mm}

\titlehead{Baltic Astronomy, vol.\,}

\titleb{SYMBIOTIC STARS: OBSERVATIONS CONFRONT THEORY}

\begin{authorl}
\authorb{J. Miko{\l}ajewska}{}
\end{authorl}

\begin{addressl}
\addressb{}{N. Copernicus Astronomical Center, Bartycka 18, 00-716 Warsaw, Poland; mikolaj@camk.edu.pl}
\end{addressl}

\submitb{Received: ; accepted: }

\begin{summary} 
In this paper, I present and discuss some recent observational results which may have important implications for our understanding of late phases of binary evolution. 
\end{summary}

\begin{keywords} stars:  binaries: symbiotic -- stars: fundamental
parameters -- stars: mass loss \end{keywords}

\resthead{Symbiotic stars: observations confront theory}
{J. Miko{\l}ajewska}

\sectionb{1}{INTRODUCTION}

Symbiotic stars are interacting binaries, consisting of an evolved giant (either a normal red giant in S-types or a Mira variable embedded in an optically thick dust shell in D-type) transferring mass to a hot and luminous white dwarf (although in a few cases a neutron star has been found). 
Since the binary must have enough room to accommodate the red giant, and in the case of D-types also its dust shell, symbiotic stars have the longest orbital periods and so the largest component separations among interacting binaries.
The interacting stars are surrouned by a rich and luminous circumstellar environment resulting from the presence of both an evolved giant with a heavy mass loss and of a hot companion abundant in ionizing photons and often emanating its own wind. 
In particular, strongly different surroundings are expected, including both ionized and neutral regions, dust forming regions, accretion and excretion disks, interacting winds and jets.  Such a complex structure makes 
symbiotic stars very important tracers of late stages of stellar evolution of low and intermediate-mass stars, and an excellent laboratory for exploring various aspects of interactions and evolution in binary systems (e.g. Corradi, Miko{\l}ajewska \& Mahoney 2003, and references therein). 
Firming links between symbiotic stars and other families of interacting binaries with evolved giants  is essential to understand, for instance, the role of such binaries in the formation of stellar jets, planetary nebulae, novae, super soft X-ray sources (SSXS), and Type Ia Supernovae (SNe Ia). Many of these are issues concerning the late stages of stellar evolution which are presently poorly known, but with important implications on our understanding of the stellar population and chemical evolution of galaxies, as well as extragalactic distance scale.

\begin{figure}[]
\vbox{
\centerline{\psfig{figure=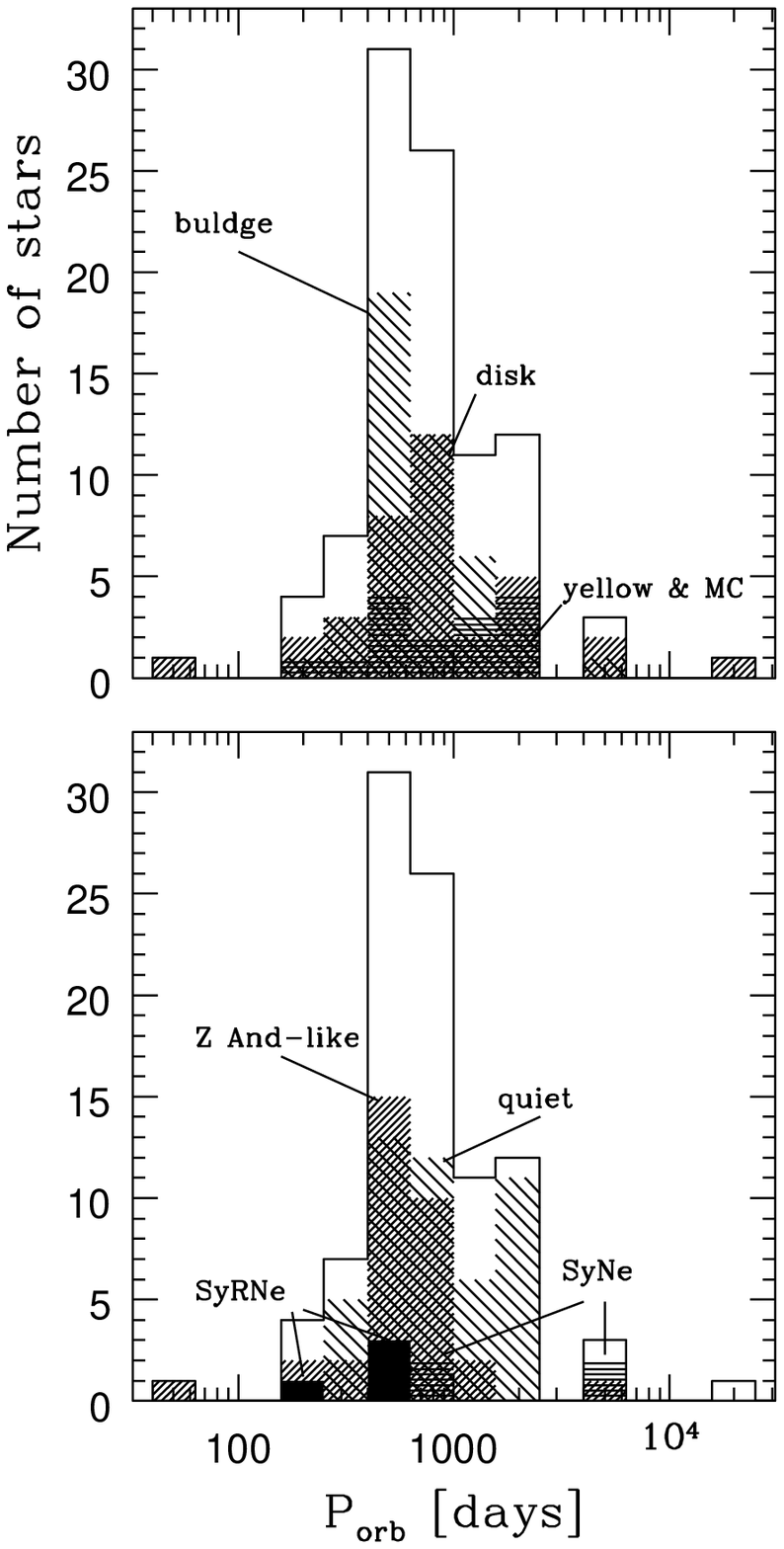,width=100mm,angle=0,clip=}}
\vspace{1mm}
\captionb{1}
{The orbital period distribution. The shaded regions denote different populations (upper panel) -- the MCs and yellow, disk and buldge, as well as different activity classes (lower panel) -- recurrent and ordinary symbiotic novae, systems with Z And-like activity, and systems without recorded outbursts.}
}
\end{figure}

Although significant progress has been done in deriving physical parameters of symbiotic stars, including orbital parameters, radii and mass loss rates (see Miko{\l}ajewska 2007, 2008, 2011a for reviews), there are important issues concerning their interactions and evolution which are not yet fully understood.

The aim of this paper is to address some problems posed by observations of symbiotic binaries to  theoretical models and binary evolution scenarios.

\sectionb{2}{ORBITAL PERIOD DISTRIBUTION}

Significant progress has been also made in deriving orbital periods using massive photometric survey databases (Gromadzki et al. 2011; Miko{\l}ajewska 2011a, and references therein). They are now known for more than 50\,\% of all S-type 
systems included in Belczy{\'n}ski et al.'s catalog (2000), although many of these periods have not yet been confirmed by radial velocity studies. Gromadzki et al. (2011) have argued that this makes $\sim 80\,\%$ of orbital periods that can be determined from photometric observations which means that the resulting period distribution should not be affected by selection effects.

Significant progress has been also made in deriving orbital periods using massive photometric survey databases (Gromadzki et al. 2011; Miko{\l}ajewska 2011a, and references therein).
They are now known for more than 50\,\% of all S-type systems included in Belczy{\'n}ski et al.'s catalog (2000), although many of these periods have not yet been confirmed by radial velocity studies. Gromadzki et al. (2011) have argued that this makes $\sim 80\,\%$ of orbital periods that can be determined from photometric observations which means that the resulting period distribution should not be affected by selection effects.

The distribution of orbital periods is shown in Fig. 1.  First of all, despite of continuously increasing number of measured periods, the main characteristics of their distribution remain practically the same as in earlier studies (e.g. Miko{\l}ajewska 2003, 2007).
In particular, they peak around 600 d, and less than 30\,\% systems have $P_{\rm orb}$ above 1000 d. 
One of extremely important issue addressed by these results is that the currently used population synthesis models (PSM) fail to reproduce the observed orbital period distribution.  The PSM predicts a $P_{\rm orb}$ distribution peaked at $\sim 1500\, \rm d$ and only 20\,\% symbiotic systems with periods below 1000 d (Lu et al. 2006).  The key to solve this problem is more advanced approach to mass transfer in symbiotic stars, and in fact in
any interacting binaries involving red giants (e.g. Podsiadlowski \& Mohamed 2007). 
In fact, there is clear evidence that in majority of S-type systems mass transfer and strong tidal interaction must have taken place before the present white dwarf was formed. Practically all systems with $P_{\rm orb}$ below $\sim 900\, \rm d$ have circular orbits, and for those with white dwarf mass above $\sim 0.5\, \rm M_{\odot}$, the present binary separation is too small and the orbital period too short to accommodate the former AGB star.

Fig. 1 shows that there is no significant difference between the orbital period distribution for the symbiotic systems in the Galactic disk and those in the buldge.  The yellow systems with a G-K giant donor and  those in the MCs may split into two populations, with $P_{\rm orb} \sim 200$--600\,d (only the low metallicity halo systems with K-type giants)  and $\geq 1000\, \rm d$ (dominated by the MC systems and D'-type systems), respectively. 

Fig. 1 also shows the $P_{\rm orb}$ distribution for different activity classes. 
It is remarkable that vast majority ($\sim 90\, \%$) of symbiotic stars with multiple outburst Z And-type activity have orbital periods below $\sim 1000\, \rm d$, whereas all four symbiotic recurrent novae (SyRNe) have orbital periods below 600 d. This may be related to higher mass transfer and accretion rates in the SyRNe and Z And-like systems than in remaining systems.
 
\sectionb{3}{MASS TRANSFER AND ACTIVITY}

\begin{figure}[]
\vbox{
\centerline{\psfig{figure=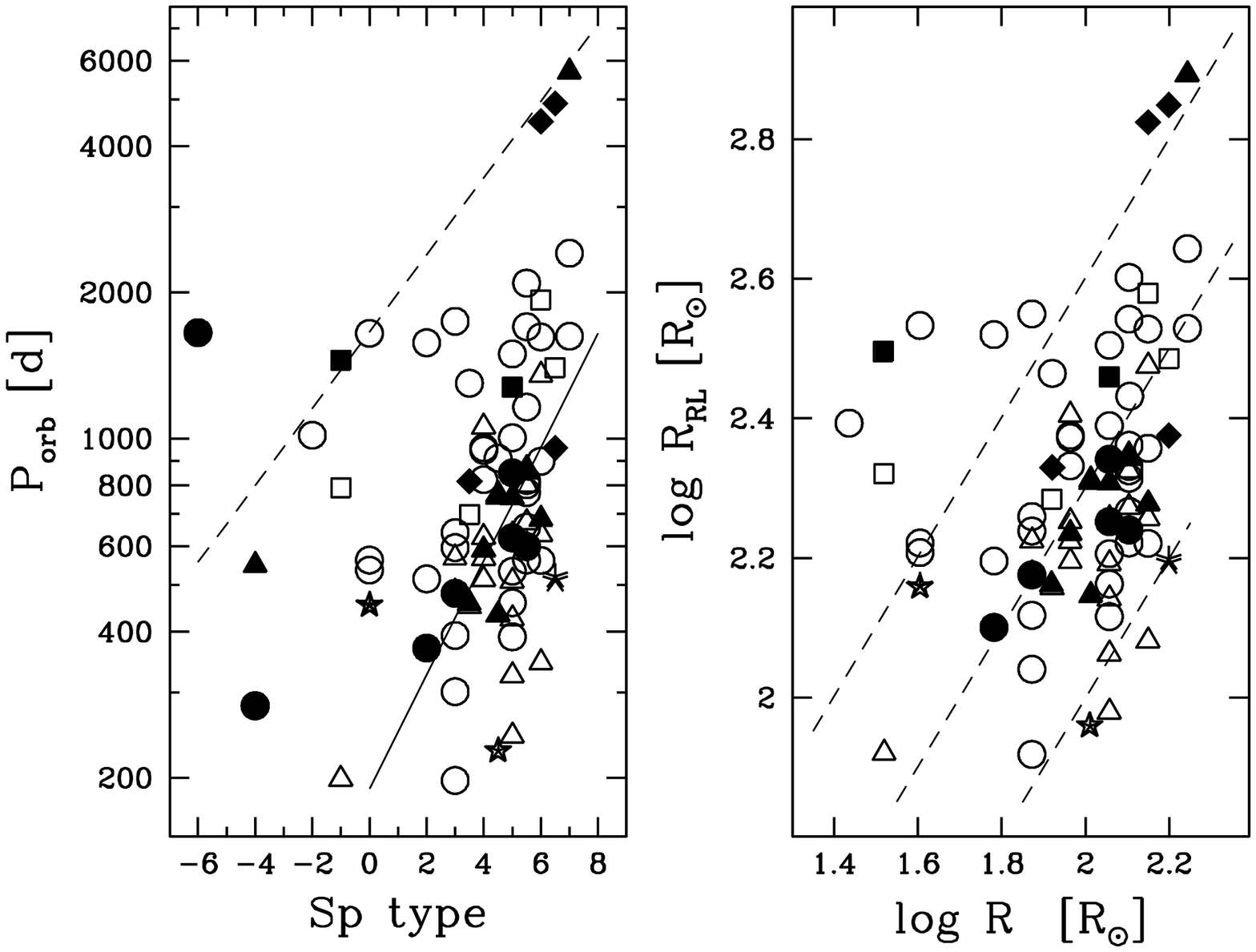,width=100mm,angle=0,clip=}}
\vspace{1mm}
\captionb{2}
{Left: Orbital periods as a function of the spectral types. Different symbols represent different activity classes: recurrent (stars) and ordinary (diamonds) symbiotic novae, 
systems with Z And-like activity (triangles), systems without recorded outbursts (circles) and systems with enhanced activity related to the periastron passage (squares). The filled symbols denote objects included in M{\"u}rset \& Schmid (1999). The lines make the period limits found by M{\"u}rset \& Schmid (1999).
Right: The Roche lobe radius versus the cool giant's radius, as derived from the spectral type, for systems with K and M-type giants. Lines denote
$R_{\rm RL}=R$, $R_{\rm RL}=2\,R$, and $R_{\rm RL}=4\,R$, respectively.}
}
\end{figure}

One of the most fundamental question concerning S-type symbiotic stars is the mode of mass transfer, via stellar wind or Roche lobe overfow (RLOF), and the possibility of an accretion disk formation.

Among the main arguments against the RLOF are the red giant radii derived from their rotational velocities, which, in most cases indicate $R_{\rm g} \sim 0.5\, R{\rm RL}$, and the relation for the minimum orbital period for symbiotic systems with red giants of a given spectral type found by M{\"u}rset \& Schmid (1999) which they interpreted as a restriction to the radius of the red giant. They also argued that this limit corresponds to a configuration where the radius of the giant reaches about half distance from its center to the inner Lagrangian point, $L_1$.
However, their relation seems to be not so obvious with the present sample of systems with known periods, a factor of 3 larger than that available for M{\"u}rset \& Schmid. The orbital periods versus the spectral types together with the period limits found by M{\"u}rset \& Schmid are plotted in the left panel of Fig. 2, whereas the right panel shows the Roche lobe radius (which, in our opinion, is more relevant that the distance to $L_1$) versus the average radius for the corresponding spectral type for systems with K and M-type giants. The average Roche lobe radius (Paczy{\'n}ski 1971) was calculated adopting the mass ratio, $M_{\rm g}/M_{\rm wd}=2.5$,  and the total mass of $2\,\rm M_{\odot}$ (Miko{\l}ajewska 2003). Many systems appear to have orbital periods significantly shorter than the lower limit of  M{\"u}rset \& Schmid, and their spectral types indicate $R_{\rm g}$ larger (and even close to) $R_{\rm RL}$. 

\begin{wrapfigure}[32]{r}[0pt]{70mm}
\vbox{
\centerline{\psfig{figure=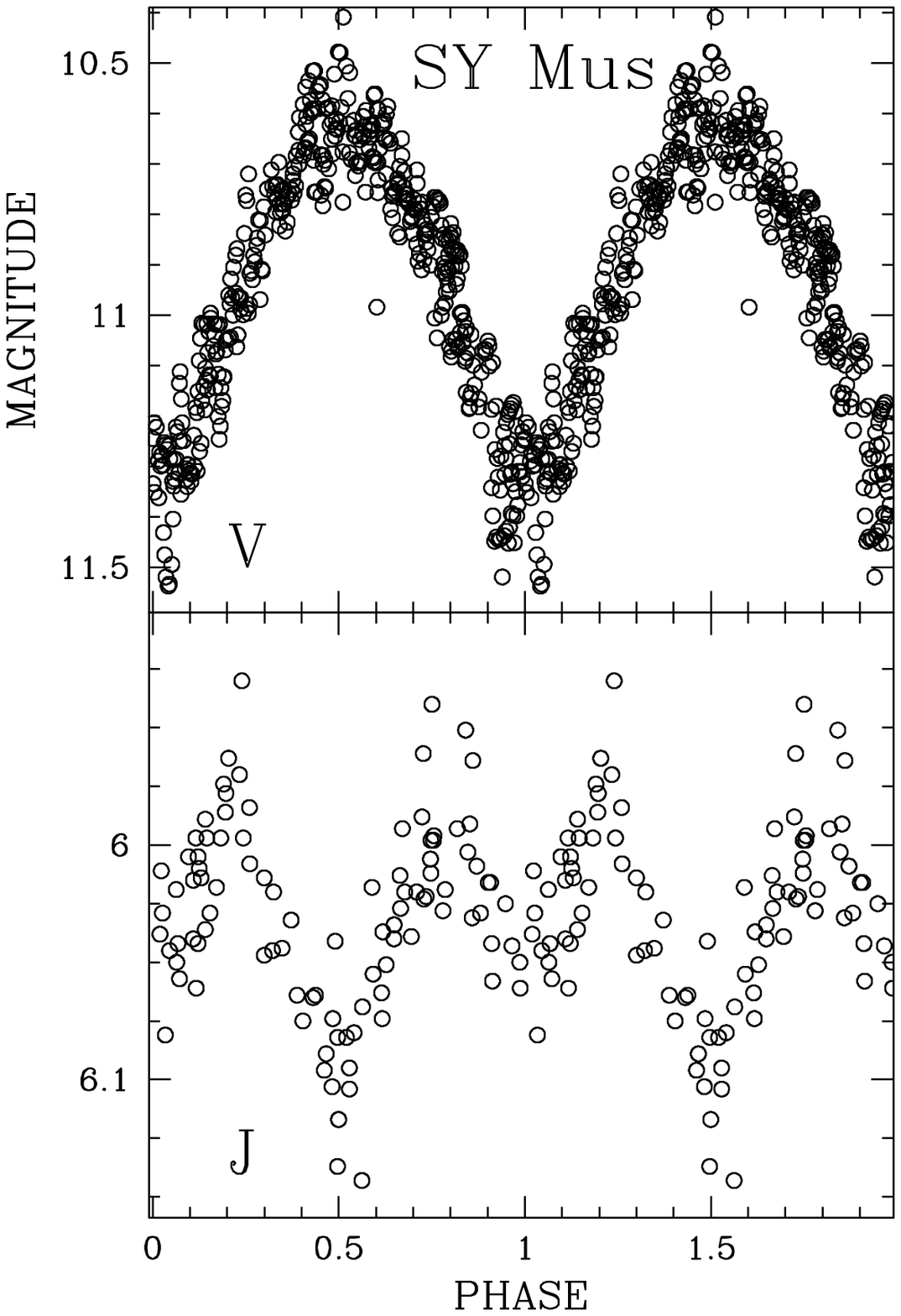,width=68truemm,clip=}}
\vspace{1mm}
\captionb{3}
{$V$ (ASAS database; Pojma{\'n}ski 2002) and $J$ (Rutkowski et al. 2007) light curves of SY Mus phased with the orbital period of 624.5 d. The ellipsoidal variability is visible only in the near infrared light.}
}
\end{wrapfigure}

Moreover, there is clear evidence for ellipsoidal light curve variations in many S-type systems with $P_{\rm orb}$ below $\sim 1000\, \rm d$, including three of the four known SyRNe (Miko{\l}ajewska 2003, 2007, 2011a; Rutkowski et al. 2007; Schaefer 2009; Yudin et al. 2005).  
Such changes can be, however, detected only in systems with relatively high inclination, and, in addition, they usually show up only in red and near infrared light (where the red giant dominates), whereas for most symbiotic stars only visual light curves are available (Fig. 3). 
This is particularly unfortunate in the case of active Z And-like systems and inactive systems with very luminous hot component like SY Mus which dominate among S-type symbiotics with shorter periods.
Anyway, it seems that ellipsoidal variability is present in $IJHK$ light curves of all multiple outburst Z And-type systems, and in some steady systems as well, with the orbital inclination above $\sim 60^{\circ}$ (Rutkowski et al. 2007). 
Such variability is, however, definitely absent in the symbiotic nova AG Peg (Rutkowski et al. 2007).

The presence of tidally distorted giants in symbiotic systems would facilitate accretion disk formation. This is particularly interesting in relation to the observed outburst characteristics of Z And-like systems which can be qualitatively accounted for by unstable accretion disk around the thermonuclear H-shell burning white dwarf (Miko{\l}ajewska 2003; Sokoloski et al. 2006) and the symbiotic recurrent novae where high accretion rates required to account for their short recurrence timescales could not be easily achieved by wind accretion. There is also continual problem with interpretation of spectra of some symbiotics, including RW Hya and SY Mus, the stable systems with tidally distorted giants. Their far ultraviolet spectral energy distributions indicate a factor of $\sim 2$ lower temperatures than emission lines which may indicate the presence of accretion disks also in these systems (e.g. Sion et al. 2002; Kolb et al. 2004; Sion et al. 2008). 

There are, however, two embarrassing problems with the ellipsoidal variability of the symbiotic giants.

First, the mass ratio, measured by radial velocity curves for systems like AR Pav, CI Cyg and BF Cyg,  is of $\sim 2$--4 (where the red giant mass donor is the more massive component), and according to the binary evolution theory they should undergo dynamically unstable mass transfer. One may argue that all these systems are about to enter unstable mass transfer, however, the very short timescales for this phase  (a few $10^3$ yrs or so) make it very unlikely.
Thus far efforts to find new mechanisms to stabilize mass transfer have not been successful (e.g. Podsiadlowski \& Mohamed 2007; Chan et al. 2010).

Second problem faced by most symbiotic systems with ellipsoidal variability is the systematic discrepancy between the radii derived from light curve analysis and those measured from rotational velocities (Miko{\l}ajewska 2007). 
The only systems which do not show such an effect are CI Cyg and AX Per.
To make the situation even more complicated, most researchers believe that the giant donor in the symbiotic recurrent nova RS Oph underfills its Roche lobe by a factor of $\sim 2$ whereas its rotational velocity is consistent with a synchronized Roche lobe-filling star (Brandi et al. 2009). 

There is no obvious solution to the 'radius--$v_{\rm rot}\,\sin i$ problem in tidally interacting binaries. Asynchronous rotation seems unlikely because of circular orbits and short timescales for the synchronization (actually shorter than that for circularization) in such binaries. 
Although it is possible that the assumption of a spherical giant with a simple limb-darkening biases the $v_{\rm rot}\,\sin i$ measurements (e.g. Orosz \& Hauschildt 2000), the observed discrepancy seems to be too large for such an explanation. 
It is also hard to believe that the ellipsoidal variations are caused by the very extended stratified atmospheres (with molecular layers) or slow stellar wind  filling the Roche lobe rather than the star itself because the effect is independent of the spectral type (from M2 on RW Hya to M6 in AR Pav), and the multicolor light curves can be fitted with single atmosphere ellipsoidal model (e.g. Yudin et al. 2005; Rutkowski et al. 2007). 
At present the most promising explanation seems to be modification of the Roche potential used for the ellipsoidal variability by including radiation effects and pulsations. For example, Schuerman (1972) analytically showed that the contact surface can shrink up to 2--3 times for stars with strong winds. Later extensions of these calculations showed that symbiotic stars can be good candidates for such an adapted Roche potential (Dermine et al. 2009).
In fact, the cool giants in most S-type systems (including SY Mus) show low amplitude pulsations (Gromadzki et al. 2007, 2011) whereas the radio observations indicate strong, $\sim 10^{-7}\, \rm M_{\odot}/yr$, winds (e.g. Miko{\l}ajewska et al. 2002).

All these possibilities can be tested by direct measurements of the radii of the red giants in brightest and closest systems by interferometric technique. 
After all, it may happen that synchronization model for binaries also needs improvement, for instance, by including stellar wind and pulsations.

\sectionb{4}{SYMBIOTIC STARS AND PROGENITORS OF TYPE IA SUPERNOVAE}

Understanding symbiotic stars may also help to solve one of the timeliest problems in modern astrophysics -- missing progenitors of SNe Ia. There is general consensus that they result from thermonuclear disruption of CO white dwarf reaching the Chandrasekhar mass either due to mass accretion from a non-degenerated companion (single-degenerate or SD model) or mass transfer between and/or merger of two white dwarfs (double-degenerate or DD model). Unfortunately, the progenitors of SNe Ia have never been observed before the explosion, and each of the proposed scenarios has its pros and cons (DiStefano \& Orio 2011, and references therein). With regard to this problem symbiotic stars are important for several reasons. First, recent observations strongly support the single-degenerate model for at least a subset of SNe Ia, in particular systems with red-giant companions resembling the S-type symbiotic RS Oph (Patat et al. 2011, and references therein). Second, as recently pointed out by DiStefano (2010), the key to solve the missing progenitor problem lies in understanding the appearance of nuclear-burning white dwarfs which according to theoretical predictions should appear as bright SSXSs whereas observations show that they are too few if them in all types of galaxies to serve as progenitors of SNe Ia, and moreover it applies to in any scenario. Whether or not symbiotic stars can be SNe Ia progenitors, most of them contain nuclear-burning white dwarfs (e.g. Miko{\l}ajewska 2003, 2011a) but the only SSXSs associated with symbiotic stars are two systems in the SMC, SMC 3 and LIN 358, and AG Dra, the galactic system with the lowest measured metallicity. 
Moreover, symbiotic stars appear to be promising candidates for both SD and DD scenarios because some systems like AR Pav, if they will pass through a common envelope phase, may become close  double white dwarf systems with total mass comparable to the Chandrasekhar limit (Miko{\l}ajewska 2011b).

\sectionb{5}{CONCLUSIONS}

Summarizing, symbiotic binaries continually pose challenges to the binary evolution theory. Understanding the mass transfer and accretion processes in these systems is not only essential for understanding their present status but also for estimating their rates as well as for {\it any} interacting binaries involving evolved giants.

\thanks{ The use of the All 
 Sky Automated Survey (ASAS) database is acknowledged.}

\References

\refb Belczy{\'n}ski, K., Miko{\l}ajewska, J., Munari, U., Ivison, R.J., Friedjung, M. 2000, A\&AS, 146, 407

\refb Brandi, E., Quiroga, C.; Miko{\l}ajewska, J.; Ferrer, O. E., Garcia, L. G. 2009, A\&A, 497, 815

\refb Chen, X., Podsiadlowski, Ph., Miko{\l}ajewska, J., Han, Z. 2010, 
in {\it V. Kalogera and M. van der Sluys, International Conference on Binaries}, AIP Conf. Proc. Ser., vol. 1314, 59

\refb  Corradi, R.L.M., Miko{\l}ajewska, J., Mahoney, T.J., eds. 2003, {\it
 Symbiotic Stars Probing Stellar Evolution}, ASP Conf. Series, Vol. 303.

\refb Dermine, T., Jorissen, A., Siess, L., Frankowski, A. 2009, A\&A, 507, 891

\refb Di Stefano, R. 2010, ApJ, 719, 474

\refb Di Stefano, R., Orio, M., eds. 2011, {\it Binary Paths to Type Ia Supernovae, IAU Symposium no 281}, Cambridge University Press, in press

\refb Gromadzki, M., Miko{\l}ajewska, J., Borawska, M., Lednicka, A. 2007, Baltic Astronomy, 16, 37

\refb Gromadzki, M., Miko{\l}ajewska, J., Soszy{\'n}ski, I. 2011, to be , submitted to Acta. Astr.

\refb Kolb, K., Miller, J., Sion, E.M., Miko{\l}ajewska, J. 2004, AJ, 128, 1790

\refb Miko{\l}ajewska, J. 2003,  in {\it Symbiotic stars probing stellar evolution}, eds. R.L.M. Corradi, J. Miko³{\l}jewska \& T.J. Mahoney, ASP Conf. Ser. vol. 303, 9 

\refb Miko{\l}ajewska, J. 2007, Baltic Astronomy, 16, 1

\refb Miko{\l}ajewska, J. 2011a, in {Physics of Accreting Compact Binaries}, ed. D. Nogami,  Universal Academy Press, Inc., in press/arXiv:1011.5657

\refb Miko{\l}ajewska, J. 2011b, in {\it Binary Paths to Type Ia Supernovae, IAU Symposium no 281}, eds. R. Di Stefano \& M. Orio, eds., Cambridge University Press, in press/arXiv:1110.1847

\refb Miko{\l}ajewska, J., Ivison, R.J., Omont, A. 2002, Adv. Space Res., 30, 2045

\refb M{\"u}rset, U., Schmid, H.M. 1999, A\&AS, 137, 473

\refb Orosz, J., Hauschildt, P.H. 2000, A\&A, 364, 2650

\refb Paczy{\'n}ski, B., 1971, ARA\&A, 9, 183

\refb Patat, F., Chugai, N. N., Podsiadlowski, Ph., Mason, E.; Melo, C., Pasquini, L. 2011, A\&A, 530, 63

\refb Podsiadlowski, Ph., Mohamed, S. 2007, Baltic Astronomy, 16, 26

\refb Pojma{\'n}ski, G.  2002, Acta Astron., 52, 397

\refb Rutkowski, A., Miko{\l}ajewska, J., Whitelock, P.A. 2007,  Baltic Astronomy, 16, 49

\refb Schaefer, B.E. 2009, ApJ, 697, 721

\refb Schuerman, D.W. 1972, Ap\&SS, 19, 351

\refb Sion, E.M., Miko{\l}ajewska, J., Bambeck, D., Dumm, T. 2002, AJ, 123, 983

\refb Sion, E.M., Miko{\l}ajewska, J., Friedjung, M. 2008, in {\it A. Evans, M.F. Bode, T.J. O'Brien, and M.J. Darnley, eds, RS Ophiuchi (2006) and the Recurrent Nova Phenomenon}, ASP Conf. ser., vol. 401, 338

\refb Sokolski, J., Kenyon, S.J., Espey, B.R. et al. 2006, ApJ, 636, 1002
 
\refb Yudin, B. F., Shenavrin, V. I., Kolotilov, E. A., Tatarnikova, A. A., Tatarnikov, A. M. 2005, Astronomy Reports, 49, 232

\end{document}